\documentclass[prc,twocolumn,showpacs]{revtex4}
\usepackage{bm}
\usepackage{amssymb}
\usepackage{amsmath}
\usepackage{graphicx}

\begin{document}
\title{Parity Violating Electron Scattering Measurements of Neutron Densities}
\author{Shufang Ban}
\author{C. J. Horowitz}\email{horowit@indiana.edu} 
\affiliation{Department of Physics and Nuclear Theory Center,
             Indiana University, Bloomington, IN 47405, USA}
\author{R. Michaels}
\affiliation{Thomas Jefferson National Accelerator Facility
Newport News, VA, USA}

\date{\today}
\begin{abstract}
Parity violating electron scattering allows model independent measurements of neutron densities that are free from most strong interaction uncertainties.   In this paper we present statistical error estimates for a variety of experiments.  The neutron radius $R_n$ can be measured in several nuclei, as long as the nuclear excited states are not too low in energy.   We present error estimates for $R_n$ measurements in $^{40}$Ca, $^{48}$Ca, $^{112}$Sn, $^{120}$Sn, $^{124}$Sn, and $^{208}$Pb.  In general, we find that the smaller the nucleus, the easier the measurement.  This is because smaller nuclei can be measured at higher momentum transfers where the parity violating asymmetry $A_{pv}$ is larger.  Also in general, the more neutron rich the isotope, the easier the measurement, because neutron rich isotopes have larger weak charges and larger $A_{pv}$.  Measuring $R_n$ in $^{48}$Ca appears very promising because it has a higher figure of merit than $^{208}$Pb.  In addition, $R_n(^{48}$Ca) may be more easily related to two nucleon and three nucleon interactions, including very interesting three neutron forces, than $R_n(^{208}$Pb).  After measuring $R_n$, one can constrain the surface thickness of the neutron density $a_n$ with a second measurement at somewhat higher momentum transfers.  We present statistical error estimates for measuring $a_n$ in $^{48}$Ca, $^{120}$Sn, and $^{208}$Pb.  Again, we find that $a_n$ is easier to measure in smaller nuclei.    
 
\end{abstract}
\smallskip
\pacs{21.10.Gv,      
25.30.Bf,
24.80.+y
 }
\maketitle

\section{Introduction}
Nuclear charge densities have been accurately measured with elastic electron scattering and have become our picture of the atomic nucleus, see for example ref. \cite{chargeden}.  These measurements have had an enormous impact.  Unfortunately, neutron densities are not directly probed in electron scattering because the neutron is uncharged \footnote{Note that there is some information on the distribution of unpaired neutrons from magnetic electron scattering.}.  Our knowledge of neutron densities comes primarily from hadron scattering experiments involving for example pions \cite{pions}, protons \cite{protons1,protons2,protons3}, or antiprotons \cite{antiprotons1,antiprotons2}.  However, the interpretation of hadron scattering experiments is model dependent because of uncertainties in the strong interactions.  Often there are uncertainties in reaction mechanism and in probe-nucleon interaction.  Particular probes may have additional uncertainties.  For example antiprotons may only probe the large radius tail of the neutron density, because of strong absorption.  Therefore neutron densities, deduced from hadron scattering, could have significant strong interaction uncertainties.

Parity violating electron scattering provides a model independent probe of neutron densities that is free from most strong interaction uncertainties.  This is because the weak charge of a neutron is much larger than that of a proton \cite{dds}.  Therefore the $Z^0$ boson, that carries the weak force, couples primarily to neutrons.  In Born approximation, the parity violating asymmetry $A_{pv}$, the fractional difference in cross sections for positive and negative helicity electrons, is proportional to the weak form factor.  This is very close to the Fourier transform of the neutron density.  Therefore the neutron density can be extracted from an electro-weak measurement \cite{dds}.  However, the Born approximation is not valid for a heavy nucleus and coulomb distortion effects must be included.  These were calculated in ref. \cite{couldist} by numerically solving the Dirac equation for an electron scattering in both the coulomb potential and a weak interaction axial vector potential.  Many details of a practical parity violating experiment to measure neutron densities have been discussed in a long paper \cite{bigprex}.    

The Lead Radius Experiment (PREX) measures the parity violating asymmetry $A_{pv}$ for 1.05 GeV electrons scattering from $^{208}$Pb at five degrees \cite{prex}.  This measurement should be sensitive to the neutron r.m.s radius of $^{208}$Pb to 1\% ($\pm0.05$ fm) and initially ran at Jefferson Laboratory in the spring of 2010.  PREX should demonstrate this new technique to measure neutron densities.  

The neutron radius of $^{208}$Pb, $R_n$, has important implications for astrophysics.  There is a strong correlation between $R_n$ and the pressure of neutron matter at densities near 0.1 fm$^{-3}$ (about 2/3 of nuclear density) \cite{alexbrown}.  The larger the pressure of neutron matter, the more neutrons are pushed out against surface tension in $^{208}$Pb, and the larger is $R_n$.  Therefore measuring $R_n$ provides very interesting information on the equation of state --- pressure as a function of density --- of neutron matter.  The equation of state is very important in astrophysics to determine the structure of neutron stars.

Recently Hebeler et al. \cite{hebeler} use chiral perturbation theory to calculate the equation of state (EOS) of neutron matter.  Their EOS has important contributions from very interesting three neutron forces.  We have detailed information on two nucleon forces from nucleon-nucleon scattering.  Furthermore, we have some information on isospin 1/2 three nucleon forces from mass 3 nuclei ($^3$He, $^3$H) and proton-deuteron scattering.  However, our experimental information on three neutron forces is limited.  Uncertainties in three neutron forces dominate the error bars of Hebeler et al.'s EOS.  From their EOS, and the above correlation of $R_n$ with pressure, they predict $R_n-R_p= 0.17 \pm 0.03$ fm.  Here $R_p$ is the known proton radius of $^{208}$Pb.  Therefore, measuring $R_n$ provides an important experimental check on fundamental neutron matter calculations, and measuring neutron densities can provide important constraints on three neutron forces.

The correlation between $R_n$ and the radius of a neutron star $r_{NS}$ is also very interesting \cite{rNSvsRn}.  Both depend on the EOS of neutron rich matter.  In general, a larger $R_n$ implies a stiffer EOS, with a larger pressure, that will also suggest $r_{NS}$ is larger.  Note that this correlation is between objects that differ in size by 18 orders of magnitude from $R_n\approx 5.5$ fm to $r_{NS}\approx 10$ km.  However, $R_n$ depends on the EOS at nuclear density and below --- some average of the surface and interior densities in $^{208}$Pb.  While, $r_{NS}$ also depends on the EOS at higher densities, since the central density of a neutron star is a few or more times nuclear density.  Therefore, measuring both $R_n$ and $r_{NS}$ determines the density dependence of the EOS.  For example if $R_n$ is relatively large then the EOS is stiff at low densities.  If at the same time $r_{NS}$ is small then the high density EOS is soft with a low pressure.  This softening of the EOS with density could strongly suggest a phase transition to a high density exotic phase such as quark matter, strange matter, or a color superconductor.   

Indeed, if one can simultaneously measure masses and radii of neutron stars with a range of masses one can determine the full neutron matter EOS, pressure as a function of density \cite{MvsREOS}.  For example, very low mass neutron stars of say 1/3 of $M_\odot$ in mass have central densities near nuclear density.  Therefore for these stars, the information in $r_{NS}$ is closely related to the information in $R_n$ \cite{lowmass}.  However these low mass neutron stars may not exist since they are difficult to form.

Recently there has been great progress in deducing $r_{NS}$ (for say one solar mass and above stars) from X-ray observations.  Ozel et al. find $r_{NS}$ is very small, below 10 km from observations of X-ray bursts \cite{ozel}.  While Steiner et al. reanalyze the burst data and include observations of neutron stars in globular star clusters \cite{steiner} .  They conclude that $r_{NS}$ is near 12 km.  From the EOS implied by these observations they predict that $R_n-R_p=0.15 \pm 0.02$ fm in $^{208}$Pb.  Note that this prediction for a nuclear size is based on astronomical observations.  If Steiner et al. is correct, the high density EOS is relatively stiff and there is little room for a significant softening due to a transition to a high density exotic phase of QCD matter.  This potentially very important Steiner et al. result can be tested by measuring $R_n$.  Finally, Demorest et al. have recently discovered a 1.97 $M_\odot$ neutron star \cite{demorest}.  This important observation rules out soft high density equations of state.

The EOS of neutron matter is closely related to the symmetry energy $S$.  This describes how the energy of symmetric nuclear matter rises as one goes away from equal numbers of neutrons and protons.   To a good approximation the energy of neutron matter is equal to the energy of symmetric nuclear matter plus the symmetry energy.  The pressure of neutron matter depends on the derivative of the energy with respect to density.  Therefore there is a strong correlation between $R_n$ and the density dependence of the symmetry energy $dS/d\rho$, with $\rho$ the baryon density.  The symmetry energy can be probed in heavy ion collisions \cite{isospindif}.  For example, $dS/d\rho$ has been extracted from isospin diffusion data \cite{isospindif2} where projectile and target nuclei with different proton to neutron ratios are collided and one observes how the neutron to proton ratio equilibrates.     
Often isospin diffusion data is analyzed with transport models, since heavy ion collisions are so complicated.  Measuring $R_n$ allows one to extract $dS/d\rho$ in a way that is independent of the complex HI dynamics.

The symmetry energy $S$ helps determine the composition of a neutron star.  Neutron stars ( assuming a nucleon phase) are thought to be about 90 \% neutrons and 10 \% protons and electrons.  The proton fraction is determined by $S$.  The larger $S$ is the larger the proton fraction.  If the proton fraction is larger than about 12 \% than a neutron $n$ near the Fermi surface can beta decay to a proton $p$ and an electron $e$ and conserve both energy and momentum.
\begin{equation}
n\rightarrow p + e + \bar\nu_e,
\end{equation}
The would be followed by $p+e\rightarrow n + \nu_e$, cooling the star by the radiation of a $\nu\bar\nu$ pair.  This is called the direct URCA process.  If $R_n$ is measured to be relatively large then $dS/d\rho$ is large, at normal density, so that $S$ is likely large at high density.  This would strongly suggest that massive neutron stars will cool rapidly by direct URCA  \cite{URCA}.  

Finally both the solid crust of a neutron star and the skin of a heavy nucleus are made of neutron rich matter at similar subnuclear densities.  If $R_n$ is large then the energy of neutron matter rises rapidly with density and this quickly favors a transition to the uniform liquid phase.  Therefore, there is a strong correlation between $R_n$ and the transition density $\rho_t$ in a neutron star from low density solid crust to liquid core \cite{cjhjp_prl}.  If $R_n$ is large then the transition density from solid crust to uniform liquid core is predicted to be low.  Many neutron star observables depend on properties of the solid crust.  Indeed, the neutron radius $R_n$ impacts a very large range of nuclear physics and astrophysics.

In this paper, we explore possible parity violating experiments, in addition to PREX, to obtain further information on neutron densities.  First, in Section \ref{sec.general} we discuss some general considerations for neutron density measurements.  Next, in Section \ref{sec.prex} we review the PREX experiment and the determination of its statistical error.  Then in Section \ref{sec.additional} we present statistical error estimates for a number of other neutron density measurements.  We conclude in Section \ref{sec.conclusions} that several measurements are feasible.  We hope that this paper will lead to discussion of the physics impacts of these additional parity violating measurements of neutron densities.   

\section{General Considerations for Parity Violating Neutron Density Experiments}
\label{sec.general}
In this section we discuss general considerations for neutron density experiments.  These considerations are necessary to control statistical and systematic errors.  First the parity violating asymmetry $A_{pv}$ is small, of order a part per million.  Therefore one needs to accumulate large statistics.  This likely requires intense beams, high electron polarization, thick targets, and large acceptance detectors.  

The need for large acceptance detectors is made more challenging by the further requirement of good energy resolution to separate out inelastically scattered electrons. This also requires a nuclear target with a relatively high first excited state.  As a result neutron density measurements on deformed heavy nuclei such as Uranium or on odd $A$ heavy nuclei may not be feasible.

A thick target, of order 10\% of a radiation length, is likely necessary.  This target will produce significant losses from bremsstrahlung, and there will be significant beam heating.  However, a high powered solid target may be simpler than a high powered liquid hydrogen target.  Furthermore, the presence of a target backing or windows may not be a significant problem.  This extra material may only slightly dilute the measured asymmetry.

The desired momentum transfer $q$ is likely set by the size of the nucleus.  The statistical error, at fixed $q$, tends to be smallest for higher beam energies and smaller scattering angles.  Thus the statistical error is often minimized by using the smallest practical scattering angle.  We define a figure of merit $FOM_x$ for using parity violation to measure a property $x$ of the neutron density, such as the r.m.s. radius $R_n$ or the surface thickness $a_n$, as follows,
\begin{equation}
FOM_x=\frac{d\sigma(\theta,E)}{d\Omega} A_{pv}(\theta,E)^2 \epsilon_x(\theta,E)^2.
\label{FOM}
\end{equation}    
Here the differential cross section is $d\sigma/d\Omega$, the parity violating asymmetry is $A_{pv}$, and $\epsilon_x$ is the sensitivity of $A_{pv}$ to changes in the quantity $x$,
\begin{equation}
\epsilon_x=\frac{d\ln A_{pv}}{d\ln x}.
\label{epsilon}
\end{equation}
If $\epsilon_x$ is large, one does not have to measure $A_{pv}$ as accurately in order to accurately determine $x$.  Often $FOM_x$ is maximized for large beam energy $E$ and small laboratory scattering angle $\theta$.  Note that $A_{pv}$ and $\epsilon_x$ depend primarily on $q$.  However they also depend somewhat on $E$ and $\theta$ separately because of Coulomb distortions.  As we calculate in the next sections, the statistical error in determining a quantity $x$ is proportional to $(FOM_x)^{-1/2}$. 

In addition to statistical errors one has to control errors from several other sources including helicity correlated beam properties, normalization, and radiative corrections.  Systematic errors from helicity correlated beam properties involve false asymmetries from beam properties that change with electron spin.  These errors may be most important for experiments measuring small asymmetries.  In general the asymmetry grows with increasing momentum transfer $q$.  Therefore, errors associated with helicity correlated beam properties may be most important for measurements at small $q$.

Normalization errors arise from imperfect knowledge of the electron beam polarization.  The Lead Radius experiment may require beam polarimetry good to one \% and a future, even more precise, experiment would likely require sub one \% polarimetry.  Finally, radiative corrections involve processes with additional real or virtual photons.  They can be divided into coulomb distortion corrections where the nucleus remains in its ground state and dispersion corrections where the nucleus is in excited intermediate states.  Coulomb distortion corrections have been accurately calculated in ref. \cite{couldist}.  Dispersion corrections have been estimated in a few cases, see for example \cite{dispersioncorrections}.  Radiative corrections will become increasingly important as the precision of a measurement is increased.

\section{The Lead Radius Experiment (PREX)}
\label{sec.prex}
\begin{figure}[ht]
\begin{center}
\includegraphics[width=3.5in,angle=0,clip=true] {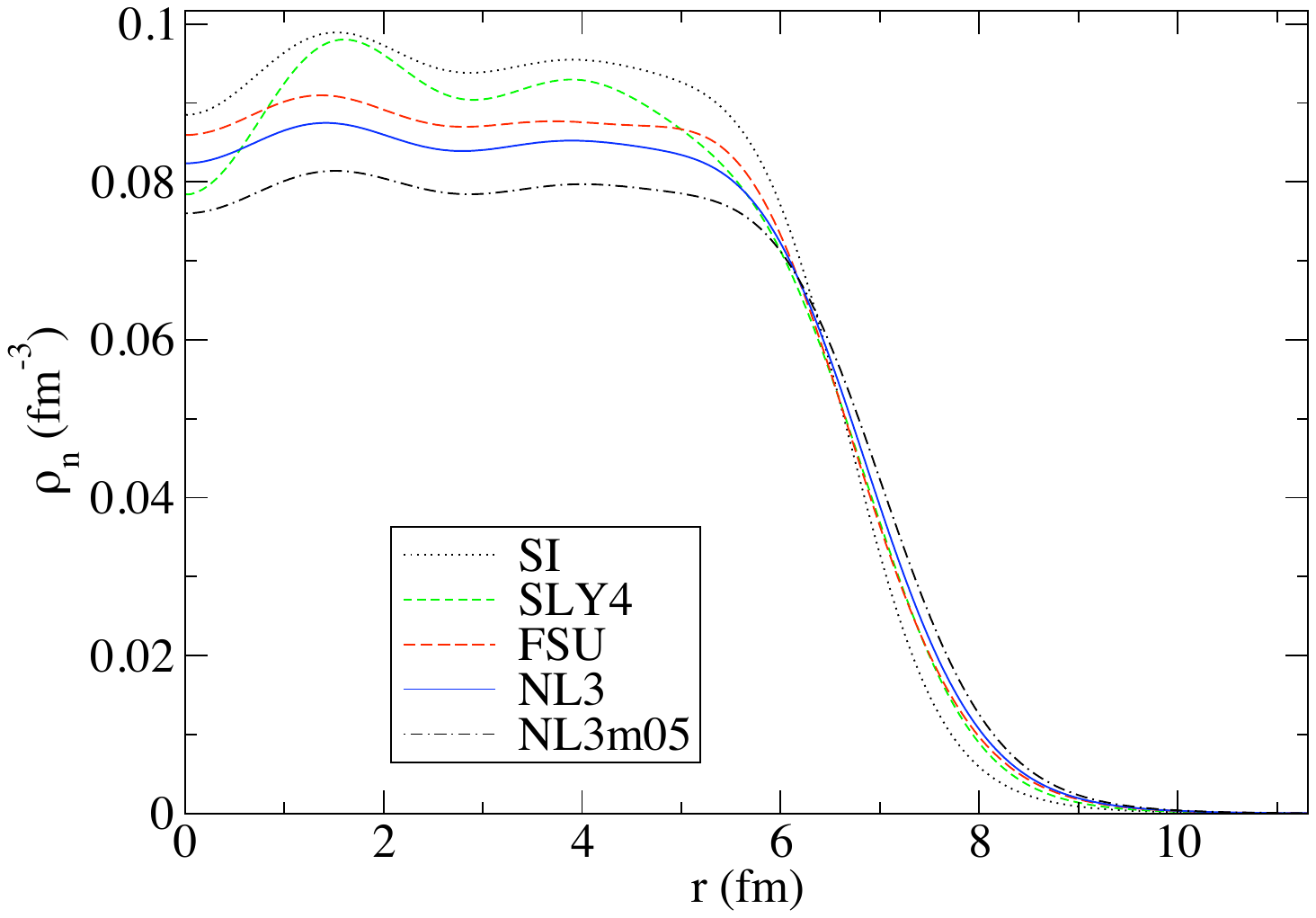}
\caption{(Color on line) Neutron density of $^{208}$Pb for the Skyrme interactions SI (black dotted curve), SLY4 (short green dashes) and relativistic mean field interactions FSUgold (longer red dashes), NL3 (solid blue), and NL3m05 (black dot dashed curve).  See Table \ref{Table1} and text.}
\label{Fig1}
\end{center}
\end{figure}

In this section we review the Lead Radius Experiment (PREX) and calculate its statistical error.  This will provide context for the discussion of additional measurements in Section \ref{sec.additional}.  PREX aims to measure the neutron r.m.s. radius $R_n$ of $^{208}$Pb with sensitivity to 1\% ($\pm 0.05$ fm).  We first collect some model neutron densities for $^{208}$Pb that are based on both nonrelativistic and relativistic mean field interactions.  These calculations yield a range of neutron radii as listed in Table \ref{Table1} and shown in Fig. \ref{Fig1}.  We modify the well known NL3 relativistic mean field interaction \cite{NL3} by adding a nonlinear coupling $\Lambda_v$ between the rho ${\bf b}^\mu$ and omega $V^\mu$ mean fields as discussed in ref. \cite{cjhjp_prl}.  The interaction Lagarangian density ${\cal L}_{\rm int}$ is,
\begin{equation}
{\cal L}_{\rm int} = \Lambda_v g_\rho^2 {\bf b}_\mu\cdot{\bf b}^\mu g_v^2 V^\mu V_\mu .
\end{equation}
The rho coupling constant $g_\rho^2$ is also modified as listed in Table \ref{Table2}.  This procedure was described in ref. \cite{cjhjp_prl} and generates neutron densities with a large range of neutron radii.  The other parameters of the NL3 interaction are unchanged \cite{cjhjp_prl} \cite{NL3}.

\begin{table}
\caption{Model root mean square proton $R_p$ and neutron $R_n$ radii for $^{208}$Pb. For NL3p06 and NL3m05 see text and Table \ref{Table2}} 
\begin{tabular}{lll}
Interaction & $R_p$ (fm) & $R_n$ (fm)\\
\toprule
Skyrme I \cite{SI} & 5.38 & 5.49 \\
Skyrme III \cite{SIII} & 5.52 & 5.65 \\
Skyrme SLY4 \cite{SLY4} & 5.46 & 5.62 \\
FSUgold \cite{FSUgold} & 5.47 & 5.68\\
NL3 \cite{NL3} & 5.46 & 5.74 \\
NL3p06 & 5.51 & 5.60 \\
NL3m05 &5.50 & 5.85 \\
\end{tabular} 
\label{Table1}
\end{table}

\begin{table}
\caption{Isovector interactions for modified NL3 \cite{NL3} parameter sets, see text and ref. \cite{cjhjp_prl}.} 
\begin{tabular}{lll}
Interaction & $\Lambda_v$ & $g_\rho^2$\\
\toprule
NL3 & 0 & 79.6 \\
NL3p06 & 0.06 & 300.\\
NL3m05 & -0.05 & 42.0 \\
\end{tabular} 
\label{Table2}
\end{table}

\begin{figure}[ht]
\begin{center}
\includegraphics[width=3.5in,angle=0,clip=true] {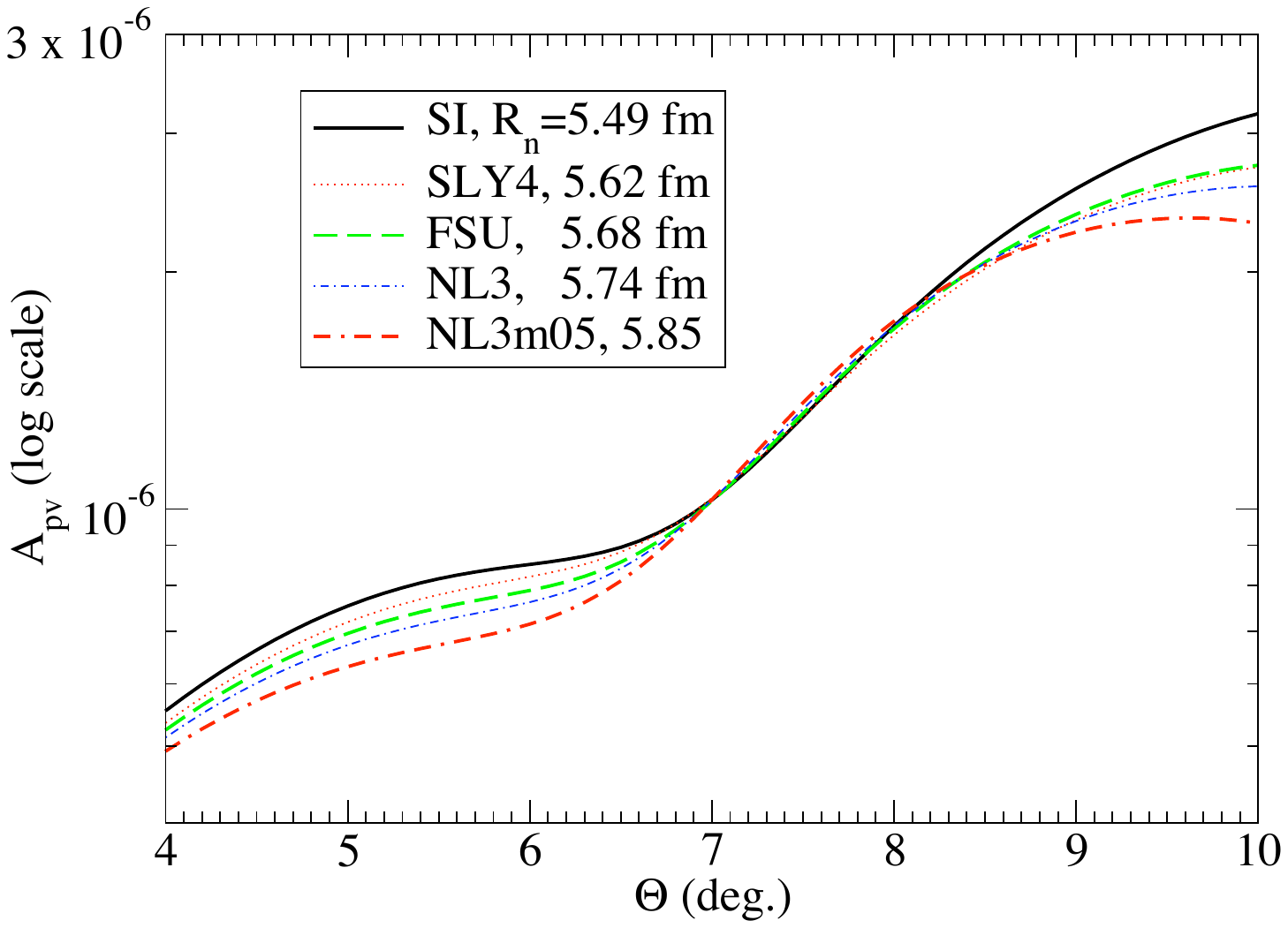}
\caption{(Color on line) Parity violating asymmetry $A_{pv}$ versus laboratory scattering angle for elastic electron scattering from $^{208}$Pb at 1.05 GeV.  The different curves are for neutron densities calculated with the indicated interactions from Table \ref{Table1}.}
\label{Fig2}
\end{center}
\end{figure}

\begin{figure}[ht]
\begin{center}
\includegraphics[width=3.5in,angle=0,clip=true] {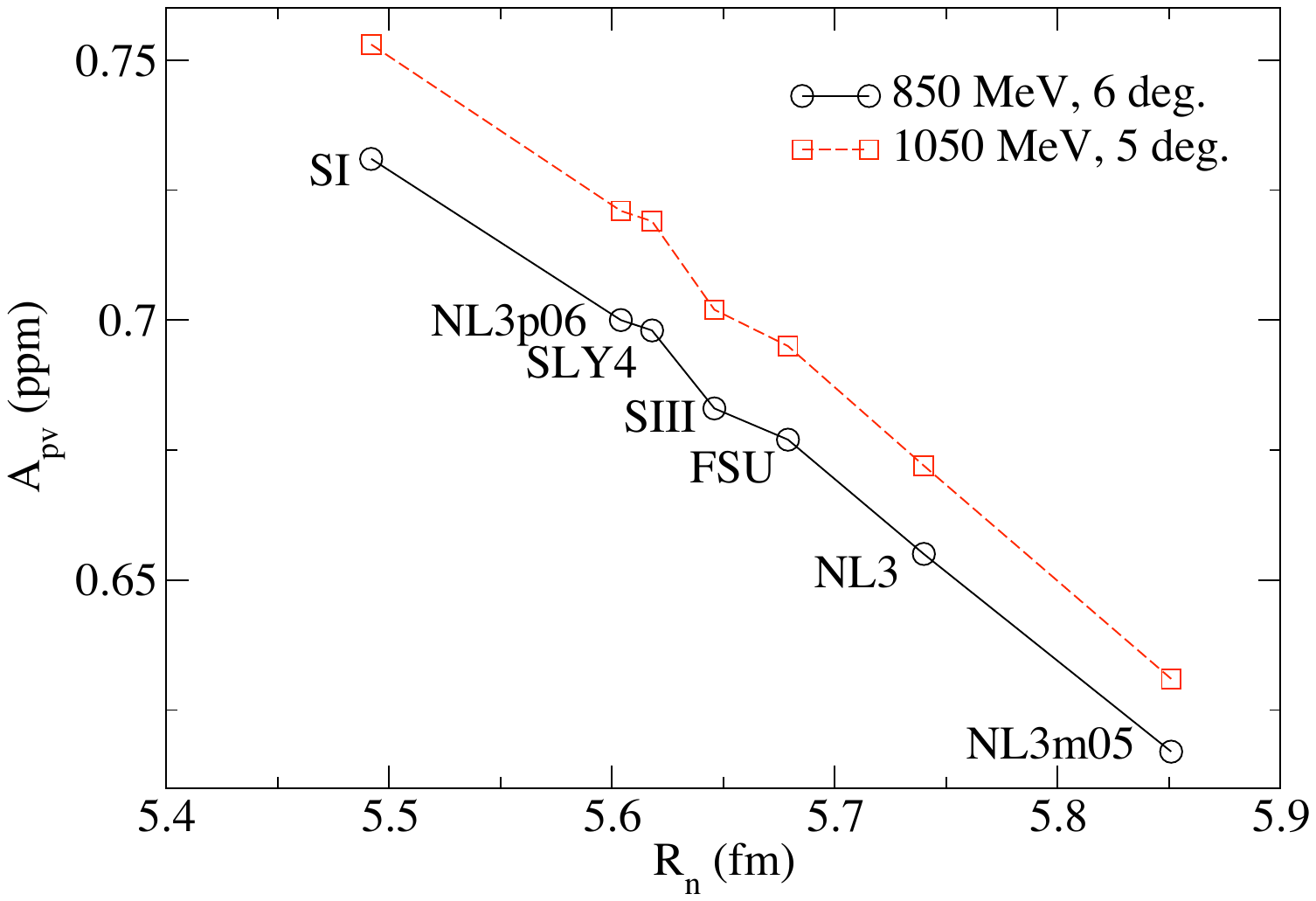}
\caption{(Color on line) Parity violating asymmetry $A_{pv}$ versus neutron r.m.s. radius $R_n$ for the model neutron densities of Table \ref{Table1}.  The red squares are for a beam energy of 1.05 GeV and a laboratory scattering angle of five degrees while the black circles are for an energy of 850 MeV and a scattering angle of six degrees.}
\label{Fig3}
\end{center}
\end{figure}

Given these neutron densities, we calculate the parity violating asymmetry $A_{pv}$ by solving the Dirac equation as discussed in ref. \cite{couldist}.  Figure \ref{Fig2} shows $A_{pv}$ versus laboratory scattering angle $\Theta$ for an electron beam energy of 1.05 GeV.  Note that we use the experimental $^{208}$Pb charge density \cite{chargeden} for these calculations instead of the model charge densities.   Figure \ref{Fig2} shows that $A_{pv}$ is very sensitive to $R_n$ for scattering angles near five degrees, and $A_{pv}$ decreases with increasing $R_n$.  In contrast, for scattering angles between seven and eight degrees, $A_{pv}$ is mostly insensitive to $R_n$ and actually increases very slightly with increasing $R_n$.  For a beam energy of 1.05 GeV and a scattering angle of five degrees, Fig. \ref{Fig3} plots $A_{pv}$ versus $R_n$ for the densities of Table \ref{Table1}.  This figure shows that there is a simple relation between $R_n$ and $A_{pv}$ so that one can deduce $R_n$ from a measured $A_{pv}$.  This will be discussed further below.

\begin{figure}[ht]
\begin{center}
\includegraphics[width=3.5in,angle=0,clip=true] {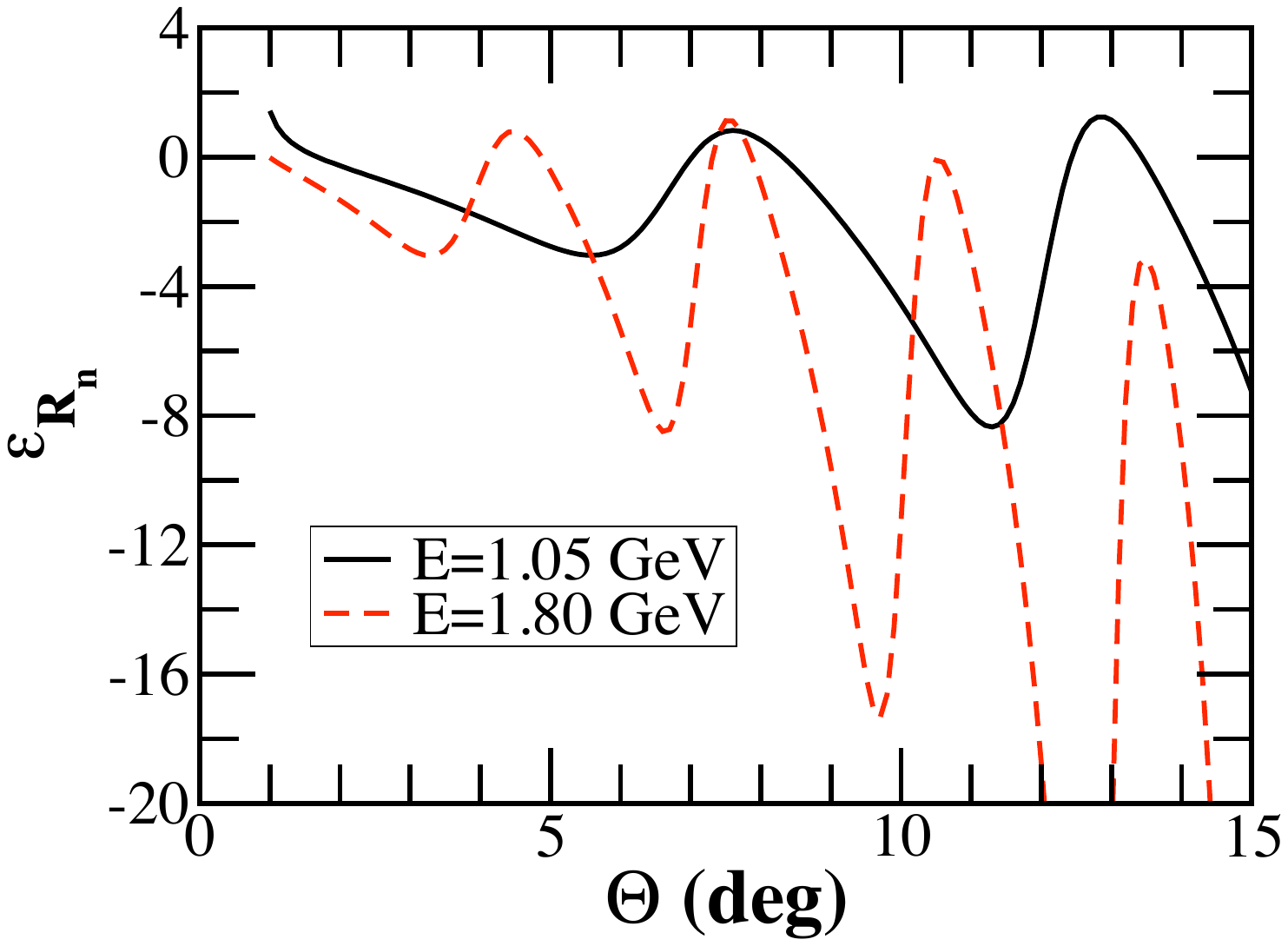}
\caption{(Color on line) Sensitivity of the parity violating asymmetry $A_{pv}$ for $^{208}$Pb to changes in the neutron radius $\varepsilon_{R_n}=\epsilon_{R_n}=d\ln A_{pv}/d\ln R_n$ versus scattering angle $\theta$ at beam energies of 1.05 GeV (solid line) and 1.8 GeV (dashed line).  The SLY4 neutron density was used.}
\label{Fig4}
\end{center}
\end{figure}

We calculate the sensitivity $\epsilon_{R_n}$ of $A_{pv}$ to the neutron radius $R_n$ as follows,
\begin{equation}
\epsilon_{R_n}=\frac{d\ln A_{pv}}{d\ln R_n}= \frac{R_n}{A_{pv}}\frac{dA_{pv}}{dR_n}.
\end{equation}
We calculate $dA_{pv}/dR_n$ by calculating the change in $A_{pv}$ when a model neutron density $\rho_n(r)$ is streched by a factor $\lambda\approx 1.01$,
\begin{equation}
\rho_n(r)\rightarrow \frac{1}{\lambda^3}\rho_n(\frac{r}{\lambda}).  
\end{equation}
Figure \ref{Fig4} shows $\epsilon_{R_n}$ versus scattering angle $\theta$ for beam energies $E$ of 1.05 and 1.8 GeV.  Near $E=1.05$ GeV and $\theta=5$ degrees $\epsilon_{R_n}$ peaks near 3.  This shows that a 3\% measurement of $A_{pv}$ is sensitive to the neutron radius to 1\% ($\pm0.05$ fm).

We now estimate statistical errors assuming the experimental parameters in Table \ref{Table3} appropriate for Hall A at Jefferson Laboratory \cite{prex} where the two High Resolution Spectrometers have solid angle acceptances near $\Delta\Omega=0.0037$ Sr each.  We somewhat arbitrarily assume a beam current $I$ of $100\mu A$.  This could be slightly optimistic.  It is easy to scale calculated count rates for other beam currents.

\begin{table}
\caption{Experimental parameters including beam current $I$, beam polarization $P$, detector solid angle $\Delta\Omega$, number of arms $N$, energy resolution $\Delta E$, and radiation loss factor $\zeta$, see text.} 
\begin{tabular}{ll}
Parameter & Value\\
\toprule
$I$ & 100$\mu A$ \\
$P$ & 0.8\\
$\Delta\Omega$ & 0.0037 Sr\\
$N$ & 2\\
$\Delta E$ & 4 MeV\\
$\zeta$ & 0.34\\
\end{tabular} 
\label{Table3}
\end{table}

We assume a target thickness of 10\% of a radiation length, see Table \ref{Table4}, and a modest energy resolution of $\Delta E=4$ MeV.  This target thickness will introduce significant radiation losses that depend somewhat on energy \cite{energyloss}.  For simplicity we multiply our rate by a radiation loss factor $\zeta=0.34$ to describe electrons that are not detected because of energy loss in the target and or bremsstrahlung during scattering.  Note that our modest energy resolution of 4 MeV will include a small contamination from the 2.6 MeV $3^-$ first excited state in $^{208}$Pb.  However this inelastic cross section is small at forward angles and one can make a crude estimate of $A_{pv}$ for this $3^-$ state \cite{bigprex}.

\begin{table}
\caption{Target thickness $t$ assuming 10\% of a radiation length.  The number of target atoms per square cm is $\rho_{\rm tar}$.} 
\begin{tabular}{lll}
Element & $t$ (cm) & $\rho_{\rm tar}$ (cm$^{-2}$)\\
\toprule
$^{208}$Pb & 0.05 & $1.6\times 10^{21}$\\
$^{120}$Sn & 0.12 & $4.4\times 10^{21}$\\
$^{48}$Ca & 1.04 & $2.4\times 10^{22}$\\
\end{tabular} 
\label{Table4}
\end{table}

The total number of electrons detected $N_{\rm total}$ during a running time $T$ is
\begin{equation}
N_{\rm total} = I T \rho_{\rm tar} \frac{d\sigma}{d\Omega}\zeta \Delta \Omega N,
\end{equation}
assuming a high efficiency detector.    Here $\rho_{tar}$ is the number density of the target in nuclei per unit area, see Table \ref{Table4}.  This allows one to be sensitive to $R_n$ with a statistical error $\Delta R_n$,
\begin{equation}
\frac{\Delta R_n}{R_n}=\Bigl( N_{\rm total} A_{\rm pv}^2 P^2 \epsilon_{R_n}^2\Bigr)^{-1/2}.
\label{eq.DRn}
\end{equation} 
Results are collected in Table \ref{Table5} for a scattering angle of 5 degrees and for a running time of $T=30$ days.  At $E=1.05$ GeV, $\Delta R_n/R_n=0.66$ \%. Alternatively, since the error scales with $T^{-1/2}$, a 1\% statistical error on $R_n$ can be obtained in $30\times (0.66)^2=13$ days.  We emphasize that this does not include many other sources of error in addition to statistics.  For example there will be an additional normalization error associated with an uncertainty $\Delta P$ in the beam polarization $P$.  This will increase the total error $\Delta R^{tot}_n/R_n$ by a factor $b$ from $\Delta R_n/R_n$ of Eq. \ref{eq.DRn},
\begin{equation}
\frac{\Delta R^{tot}_n}{R_n}=b\frac{\Delta R_n}{R_n},
\end{equation}
with $b$,
\begin{equation}
b=\Bigl[ 1+\frac{\bigl( \frac{\Delta P}{P}\bigr)^2}{\bigl(\frac{\Delta R_n\epsilon_{R_n}}{R_n}\bigr)^2} \Bigr]^\frac{1}{2}.
\end{equation}
At present $P$ can be determined with an error $\Delta P$ of order 1\% to 2\%. 

This estimate of 13 days for a 1\% determination of  $R_n$ may be slightly optimistic compared to the actual PREX experiment because the experimental acceptance includes a range of scattering angles from $\approx 4.5$ to 7 degrees and the angle averaged $A_{pv}$ may be slightly less sensitive to $R_n$ than $\epsilon_{R_n}$ at 5 degrees.  Furthermore the actual beam current could be less than 100 $\mu A$.  Nevertheless, our estimate of 13 days for a 1\% measurement of $R_n$ in $^{208}$Pb provides a benchmark that is based on our assumptions.  In Section \ref{sec.additional} we will compare this to the running time for some other neutron density measurements. 

Note that the results in Table \ref{Table5} for $^{208}$Pb at larger energies than 1.05 GeV assume a given shape for the neutron density.  At higher energies one is also increasingly sensitive to the surface thickness, see below, and other features of the neutron density in addition to the radius.  Therefore one needs to interpret the sensitivity to $R_n$ given in Table \ref{Table5} with care.

\begin{table}[h]
 \centering
 \caption{Statistical error estimates for measuring $R_n$ in 30 days.   Results are first presented for $^{208}$Pb, $^{48}$Ca, and $^{40}$Ca at a laboratory scattering angle of five degrees and then $^{48}$Ca results are also presented for a scattering angle of four degrees, see text. The neutron and proton densities are calculated in the Skyrme HF theory with the SLY4 interaction.}

\vspace{0.0em}
 \begin{tabular}{c|c|cc|cccc}
 \hline \hline
  \multicolumn{1}{c}{Nucleus}
 &\multicolumn{1}{|c}{E}
 &\multicolumn{1}{|c}{$A_{pv}(5^o)$}
 &\multicolumn{1}{c}{$\frac{d\sigma}{d\Omega}(5^o)$}
 &\multicolumn{1}{|c}{Rate$(5^o)$}
 &\multicolumn{1}{c}{$\epsilon_{R_n} $}
 &\multicolumn{1}{c}{$\Delta R_n/R_n$} \\
         &GeV&ppm&mb/str&MHz/arm&&\%\\
  \hline
  $^{208}$Pb& 1.05 & 0.7188 & 1339 & 1736
            & -2.762 & 0.6637 \\
  \hline
 $^{48}$Ca   & 1.80 & 2.358 & 8.630 & 164.3 & -4.266 & 0.4258 \\
 \hline
    $^{40}$Ca & 1.90& 2.301& 5.832 & 111.0 &-3.920& 0.5777\\
 \hline\hline
      &$E$&$A_{pv}(4^o)$&$\frac{d\sigma}{d\Omega}(4^o)$&Rate$(4^o)$&$\epsilon_{R_n} (4^o$)
      &$\Delta R_n/R_n$\\\hline
 $^{48}$Ca & 2.20 & 2.290 & 16.56 & 315.2 & -3.961 & 0.3409\\
 \hline\hline
 \end{tabular}
 \label{Table5}
 \end{table}

\section{Additional neutron density measurements}
\label{sec.additional}
In this section we consider other parity violating neutron density measurements.  First, in Subsection \ref{subsec.an208Pb} we explore a measurement of the surface thickness $a_n$ of the neutron density in $^{208}$Pb.  We assume this measurement is after the neutron radius $R_n$ has been measured.  In Subsection \ref{subsec.48Ca} we look at $R_n$ and $a_n$ measurements for $^{48}$Ca, while in Subsection \ref{subsec.120Sn} we consider the neutron radius in Tin isotopes $^{112}$Sn, $^{120}$Sn, and $^{124}$Sn.

\subsection{Surface thickness $a_n$ in $^{208}$Pb}
\label{subsec.an208Pb}
\begin{figure}[ht]
\begin{center}
\includegraphics[width=3.5in,angle=0,clip=true] {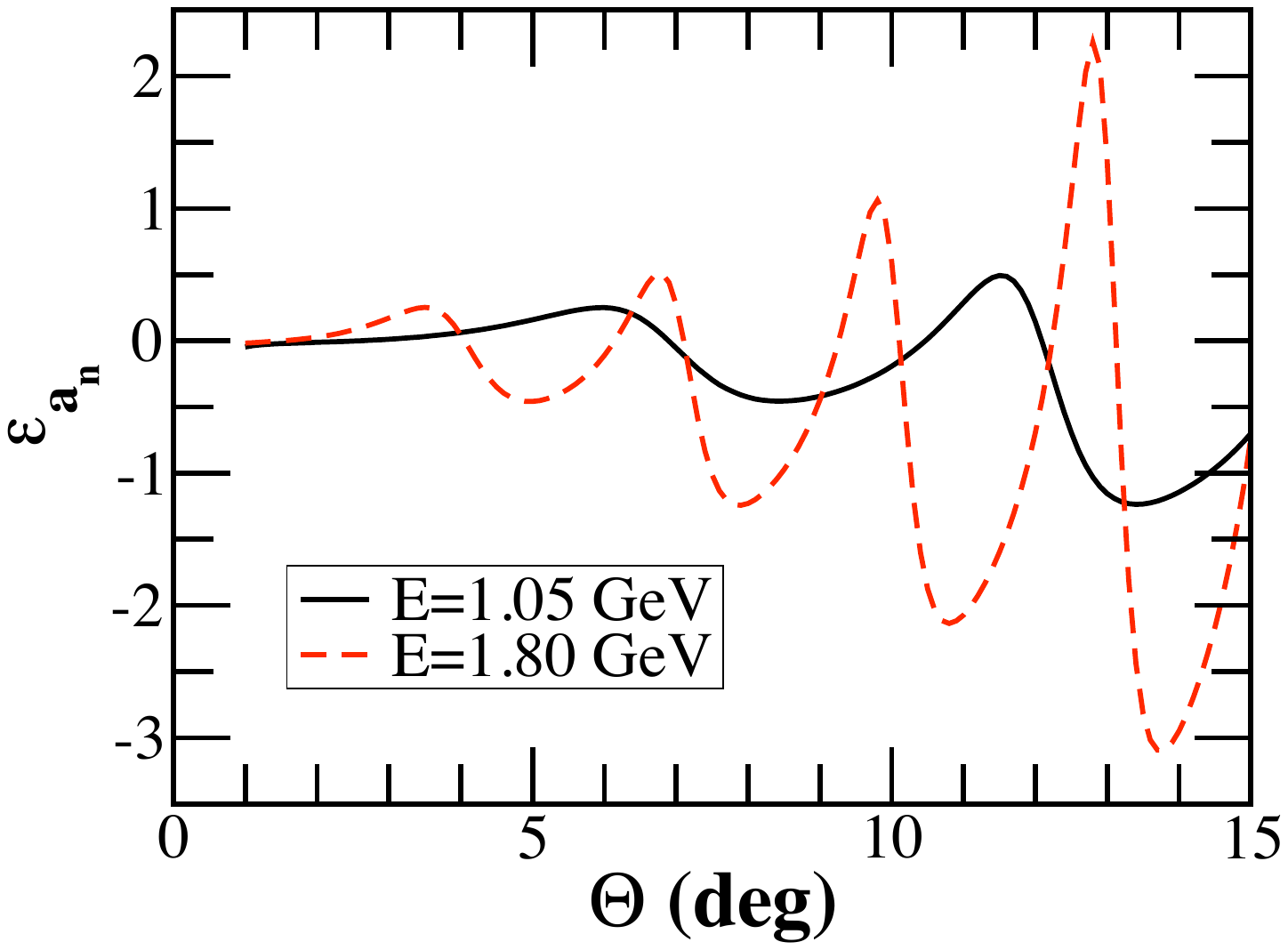}
\caption{(Color on line) Sensitivity of the parity violating asymmetry $A_{pv}$ for $^{208}$Pb to changes in the neutron surface thickness $a_n$, see Eq. \ref{woodsaxon}, where $\varepsilon_{a_n}=\epsilon_{a_n}=d\ln A_{pv}/d\ln a_n$ versus scattering angle $\theta$ at beam energies of 1.05 GeV (solid line) and 1.8 GeV (dashed line).  The SLY4 neutron density was used.}
\label{Fig5}
\end{center}
\end{figure}

To explore a surface thickness measurement of the neutron density in $^{208}$Pb we model the neutron density with a Wood Saxon form,
\begin{equation}
\rho_n(r)=\rho_0/[1+\exp(r-R_0)/a_n].
\label{woodsaxon}
\end{equation}
An approximate fit of Eq. \ref{woodsaxon} to the SLY4 neutron density yields $a_n\approx 0.55$ fm.  We calculate $A_{pv}$ for this neutron density and the sensitivity,
\begin{equation}
\epsilon_{a_n}=\frac{d\ln A_{pv}}{d\ln a_n}
\label{epsilonan}
\end{equation}
to small changes in $a_n$.  We assume that the neutron r.m.s. radius $R_n$ has been fixed by an earlier measurement.  Therefore we calculate the derivative in Eq. \ref{epsilonan} at fixed $R_n$ and not at fixed $R_0$ (the parameter in Eq. \ref{woodsaxon}).  Indeed as $a_n$ is changed $R_0$ is also changed so that $R_n$ remains fixed.  Figure \ref{Fig5} shows $\epsilon_{a_n}$ for beam energies of 1.05 and 1.8 GeV.  The statistical sensitivity of a measurement of $a_n$ is
\begin{equation}
\frac{\Delta a_n}{a_n} = \Bigl(N_{\rm total} A_{pv}^2 P^2 \epsilon_{a_n}^2\Bigr)^{-1/2}.
\label{deltaan}
\end{equation}
Results for $\Delta a_n$ are collected in Table \ref{Table6}.  In 30 days of running at 1.8 GeV and 5 degrees one can obtain a statistical error of 7.9 \%, using the experimental parameters in Table \ref{Table3}.  We emphasize that this is a sensitivity to $a_n$ of 7.9 \% given our assumptions that $R_n$ is precisely known and that the neutron density has a Wood Saxon form.

\begin{table}[h]
 \centering
 \caption{Statistical error estimates for measuring $a_n$ in 30 days
 for nuclei $^{208}$Pb, assuming $a_n\approx 0.55$ fm, and $^{48}$Ca, assuming $a_n\approx 0.50$ fm.
 The proton density is calculated in the Skyrme HF theory with
 SLY4 and the neutron density is of Wood-Saxon form,
 $\rho(r)=\rho_0/[1+exp(\frac{r-R_0}{a_n})]$.
 }
 \label{Table6}
\vspace{0.0em}
 \begin{tabular}{c|c|cc|cccc}
 \hline \hline
  \multicolumn{1}{c}{Nucleus}
 &\multicolumn{1}{|c}{E}
 &\multicolumn{1}{|c}{$A_{pv}(5^o)$}
 &\multicolumn{1}{c}{$\frac{d\sigma}{d\Omega}(5^o)$}
 &\multicolumn{1}{|c}{Rate$(5^o)$}
 &\multicolumn{1}{c}{$\epsilon_{a_n} (5^o)$}
 &\multicolumn{1}{c}{$\Delta a_n/a_n$} \\
         &GeV&ppm&mb/str&MHz/arm&&\%\\
  \hline
$^{208}$Pb & 1.80 & 2.085 & 40.59 & 52.61 &-0.4582 & 7.924 \\
     \hline
$^{48}$Ca    & 2.15 & 2.744 & 1.103 & 21.00 & 1.474 & 2.962 \\
  \hline \hline
 \end{tabular}
 \end{table}

\subsection{Neutron Densities of $^{48}$Ca and $^{40}$Ca}
\label{subsec.48Ca}
The neutron density of $^{48}$Ca is very interesting.  First $^{48}$Ca is one of the lightest neutron rich closed shell nuclei, and is much lighter than $^{208}$Pb.  Therefore $R_n(^{48}$Ca) may provide independent information that is not contained in $R_n(^{208}$Pb).  In addition, more microscopic calculations, using for example coupled cluster or no core shell model approaches, may soon be feasible for $^{48}$Ca.  These are presently not feasible for $^{208}$Pb.  This may allow one to relate neutron density measurements in $^{48}$Ca more closely to microscopic nucleon-nucleon and three nucleon forces.  In particular, one would like to probe poorly known three neutron forces with neutron density measurements.   Furthermore, $^{48}$Ca is a relatively light double beta decay nucleus.  This further motivates nuclear structure studies that could also probe its neutron density.

Finally, in addition to $^{48}$Ca, the isotope $^{40}$Ca is a stable $N=Z$ nucleus where it is expected that $R_n$ should be close to, but slightly smaller than, $R_p$.  Therefore one can compare a variety of measurements on the two Ca isotopes.  Indeed, a parity violating measurement of $R_n$ for $^{40}$Ca provides a fundamental check on the whole procedure to use parity violating electron scattering to measure neutron densities.  Many systematic errors will cancel in combining two parity violating measurements to determine $R_n(^{48}$Ca$) - R_n(^{40}$Ca).  For example, radiative corrections are expected to be very similar for the two isotopes.  

\begin{figure}[ht]
\begin{center}
\includegraphics[width=3.5in,angle=0,clip=true] {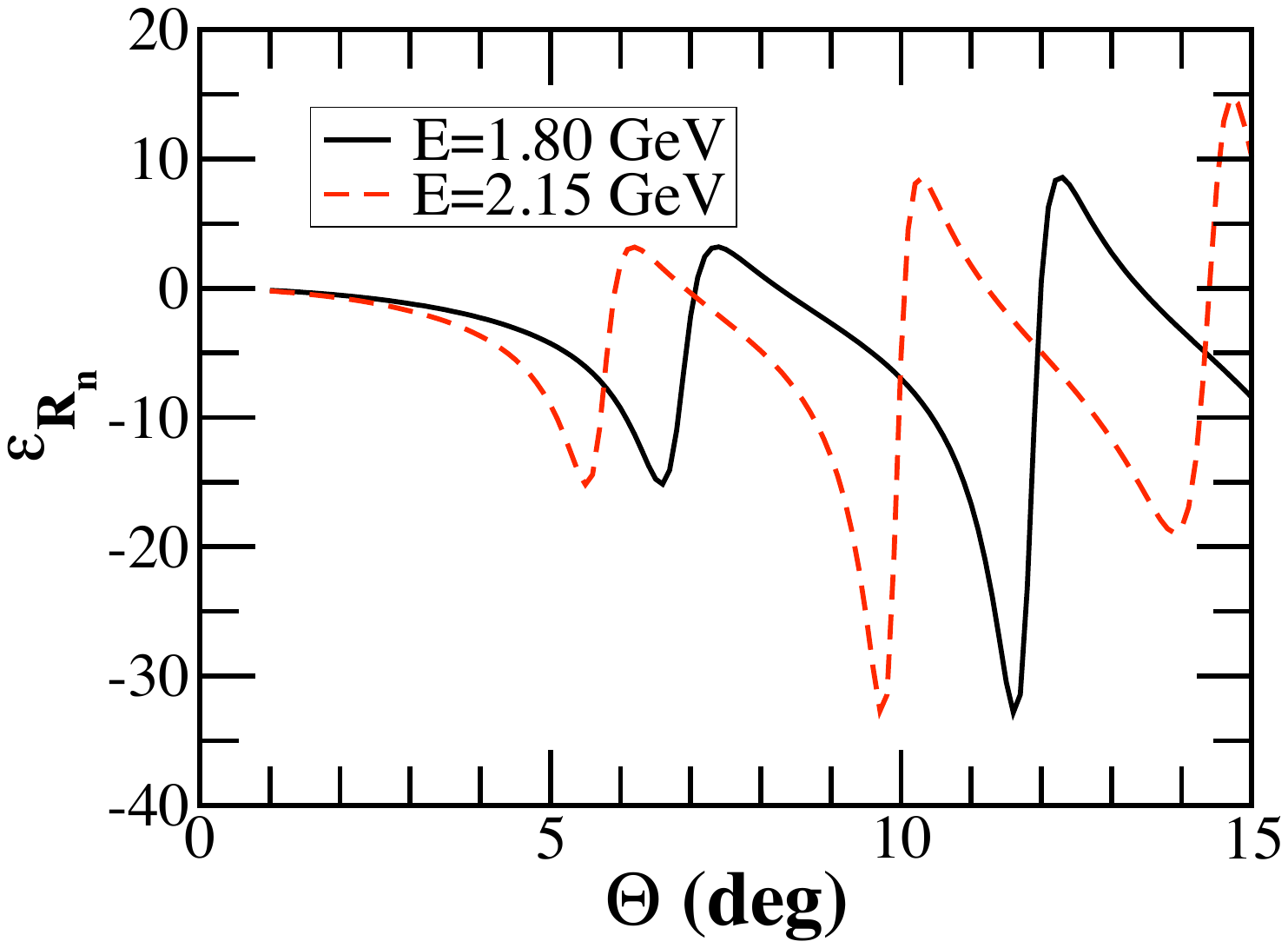}
\caption{(Color on line) Sensitivity of the parity violating asymmetry $A_{pv}$ for $^{48}$Ca to changes in the neutron radius $\varepsilon_{R_n}=\epsilon_{R_n}=d\ln A_{pv}/d\ln R_n$ versus scattering angle $\theta$ at beam energies of 1.8 GeV (solid line) and 2.15 GeV (dashed line).  The SLY4 neutron density was used.}
\label{Fig6}
\end{center}
\end{figure}

The sensitivity of $A_{pv}$ to $R_n$ for $^{48}$Ca is shown in Fig. \ref{Fig6}.  At $E=1.8$ GeV and $\theta=5$ degrees one can measure $R_n$ with a statistical sensitivity of 0.43 \% in 30 days, see Table \ref{Table5}.  Alternatively, one can measure $R_n$ with a statistical error of 1\% in only 5.5 days.  This is less than half of the beam time required for a 1\% measurement in $^{208}$Pb, as we discuss below.   Table \ref{Table5} also shows that $R_n$ in $^{40}$Ca can be measured at $E=1.90$ GeV and $\theta=5$ degrees to 0.6\% in 30 days, or a 1\% error is possible in 11 days.  Calcium 40 is slightly harder to measure than $^{48}$Ca because $A_{pv}$ is smaller, given the smaller weak charge, and because the sensitivity $\epsilon_{R_n}$ is slightly smaller.

It is interesting to compare $^{48}$Ca to $^{208}$Pb.  First, Table \ref{Table5} shows that the optimal energy, and momentum transfer $q$, for a measurement in $^{48}$Ca is higher than in Pb.  This immediately follows because $R_n$ in Ca is smaller than lead so the product $q R_n$ stays approximately constant.  This higher $q$ insures that $A_{pv}$ is larger for the Ca measurement, since $A_{pv}$ scales approximately with $q^2$.   Finally the Ca figure of merit is larger than that for Pb because of the higher $A_{pv}$ and because $\epsilon_{R_n}$ is slightly larger.

All of the above results assume a laboratory scattering angle $\theta=5$ degrees.  The existing PREX septum bends 1.05 GeV electrons, scattered at 5 degrees, into the High Resolution Spectrometers.  Note that this septum may not have a strong enough magnetic field to work at higher energies.  With the planned energy upgrade at Jefferson Laboratory, 2.2 GeV may be a good energy for a neutron density measurement.  This would allow a single pass beam to be used for the measurement while, at the same time, higher energy multi-pass beams are sent to other experiments.  A 2.2 GeV neutron radius measurement on $^{208}$Pb may need detectors at very small angles just over two degrees.  However a 2.2 GeV $R_n$ measurement on $^{48}$Ca only needs detectors near four degrees.  Therefore Table \ref{Table5} also lists results for $^{48}$Ca at four degrees.  Because of the high beam energy the figure of merit is very good.  In 30 days one could, in principle, reach a sensitivity of 0.34 \%.  Alternatively, one only needs 3.5 days to get 1 \% statistics for $R_n$.   This is a factor of four shorter time than for PREX.  If a new septum could be designed to work at four degrees and 2.2 GeV, this would allow a large improvement in the figure of merit over PREX.  Furthermore, many systematic errors depend on the absolute size of $A_{pv}$.  These may be easier to deal with in a $^{48}$Ca measurement, than for PREX, because $A_{pv}$ is larger.

\begin{figure}[ht]
\begin{center}
\includegraphics[width=3.5in,angle=0,clip=true] {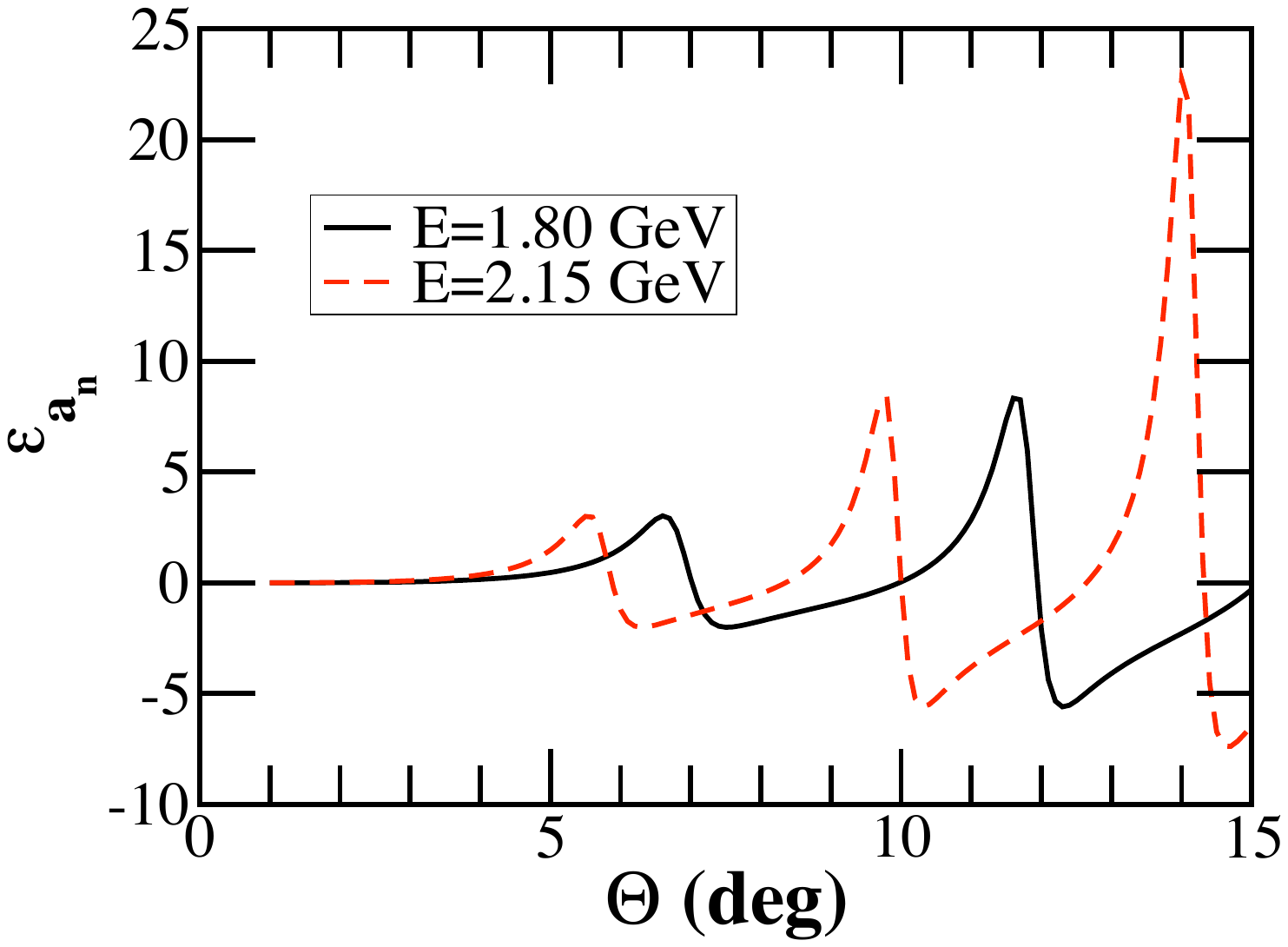}
\caption{(Color on line) Sensitivity of the parity violating asymmetry $A_{pv}$ for $^{48}$Ca to changes in the neutron surface thickness $a_n$, see Eq. \ref{woodsaxon}, $\varepsilon_{a_n}=\epsilon_{a_n}=d\ln A_{pv}/d\ln a_n$ versus scattering angle $\theta$ at beam energies of 1.8 GeV (solid line) and 2.15 GeV (dashed line).}
\label{Fig7}
\end{center}
\end{figure}

The sensitivity of $A_{pv}$ to the surface thickness $a_n$ for $^{48}$Ca is shown in Fig. \ref{Fig7}.  At $E=2.15$ GeV and $\theta=5$ degrees, one is sensitive to $a_n$ with a statistical error of 3.0\% after 30 days, see Table \ref{Table6}.  This is much smaller than the 7.9\% error for $a_n$ in $^{208}$Pb.  There are two reasons for this dramatically increased sensitivity.  First, $A_{pv}$ is large, 2.7 ppm for $^{48}$Ca at 2.15 GeV.  Second, one is much more sensitive to $a_n$ in $^{48}$Ca than in $^{208}$Pb because $^{48}$Ca is mostly surface while the surface is only a small part of $^{208}$Pb.  This leads to a much larger $\epsilon_{a_n}$.

\subsection{Neutron Densities of $^{112}$Sn, $^{120}$Sn, and $^{124}$Sn}
\label{subsec.120Sn}
The neutron density of Tin isotopes are interesting for a number of reasons.  First a variety of Tin isotopes are available for experiments with either stable or radioactive beams.  Indeed heavy ion collisions with Tin isotopes have been used to probe the density dependence of the symmetry energy \cite{heavyioncollisions}.  The neutron radius of a heavy nucleus has been shown to be closely related to the density dependence of the symmetry energy.  A rapid density dependence implies a large pressure for neutron rich matter and this give a large neutron radius\cite{cjhjp_prl}.  Second, pairing corrections may be larger in Tin isotopes than in a closed shell nucleus such as $^{208}$Pb, and these corrections could impact $R_n$.  

Table \ref{Table7} presents statistical error estimates for $R_n$ measurements in $^{112}$Sn, $^{120}$Sn, and $^{124}$Sn.   In general there is a smooth dependence on neutron number $N$ with larger $N$ isotopes being somewhat easier to measure because they have larger weak charges and $A_{pv}$.   We expect similar results for other even $N$ Sn isotopes.   We see that $R_n$ can be measured in $^{120}$Sn at $E=1.25$ GeV and $\theta=5$ degrees with a statistical sensitivity of 0.56 \% after 30 days.  This is a slightly smaller error than for $^{208}$Pb.  Although $^{120}$Sn has a smaller cross section than $^{208}$Pb, the asymmetry $A_{pv}$ is larger for $^{120}$Sn than for $^{208}$Pb and this more than compensates for the smaller cross section.     Finally Table \ref{Table8} presents statistical error estimates for measuring the surface thickness $a_n$ in $^{120}$Sn.  This could be measured at $E=2.15$ GeV and $\theta=5$ degrees with a statistical sensitivity of 5.1\% after 30 days.  Again this is a smaller error than for $^{208}$Pb.  Finally, we expect similar results for an $a_n$ measurement in other even $N$ Sn isotopes.

\begin{table}[h]
 \centering
 \caption{Statistical error estimates for measuring $R_n$ for $^{112}$Sn, $^{120}$Sn, and $^{124}$Sn in 30 days.   The neutron and proton densities are calculated in the Skyrme HF theory with the SLY4 interaction.}
 \label{Table7}
\vspace{0.0em}
 \begin{tabular}{c|c|cc|cccc}
 \hline \hline
  \multicolumn{1}{c}{Nucleus}
 &\multicolumn{1}{|c}{E}
 &\multicolumn{1}{|c}{$A_{pv}(5^o)$}
 &\multicolumn{1}{c}{$\frac{d\sigma}{d\Omega}(5^o)$}
 &\multicolumn{1}{|c}{Rate$(5^o)$}
 &\multicolumn{1}{c}{$\epsilon_{R_n} (5^o$)}
 &\multicolumn{1}{c}{$\Delta R_n/R_n$} \\
         &GeV&ppm&mb/str&MHz/arm&&\%\\
  \hline
$^{112}$Sn  & 1.30 & 1.099 & 187.2 & 654.7 & -3.157 & 0.6183 \\

  \hline
$^{120}$Sn & 1.25 & 1.124 & 230.9 & 807.8 & -3.070 & 0.5599 \\
  \hline
$^{124}$Sn   & 1.25 & 1.160 & 223.4 & 781.4 & -3.172 & 0.5337 \\
 \hline \hline
 \end{tabular}
 \end{table}

 \begin{table}[h]
 \centering
 \caption{Statistical error estimates for measuring $a_n$ in 30 days
 for $^{120}$Sn, assuming $a_n\approx 0.55$ fm.
 The proton density is calculated in the Skyrme HF theory with
 SLY4 and the neutron density is of Wood-Saxon form,
 $\rho(r)=\rho_0/[1+exp(\frac{r-R_0}{a_n})]$.
 }
 \label{Table8}
\vspace{0.0em}
 \begin{tabular}{c|c|cc|cccc}
 \hline \hline
  \multicolumn{1}{c}{Nucleus}
 &\multicolumn{1}{|c}{E}
 &\multicolumn{1}{|c}{$A_{pv}(5^o)$}
 &\multicolumn{1}{c}{$\frac{d\sigma}{d\Omega}(5^o)$}
 &\multicolumn{1}{|c}{Rate$(5^o)$}
 &\multicolumn{1}{c}{$\epsilon_{a_n} (5^o)$}
 &\multicolumn{1}{c}{$\Delta a_n/a_n$} \\
         &GeV&ppm&mb/str&MHz/arm&&\%\\
  \hline
$^{120}$Sn  & 2.15 & 3.044 & 4.470 & 15.63 & -0.8889 & 5.131 \\
 \hline \hline
 \end{tabular}
 \end{table}

\section{Discussion and conclusions}
\label{sec.conclusions}
Parity violating electron scattering allows model independent measurements of neutron densities that are free from most strong interaction uncertainties.   In this paper we present statistical error estimates for a variety of experiments.  The neutron radius $R_n$ can be measured in several nuclei, as long as the nuclear excited states are not too low in energy.   In general, we find that the smaller the nucleus, the easier the measurement.  This is because smaller nuclei can be measured at higher momentum transfers where the parity violating asymmetry $A_{pv}$ is larger.  Also in general, the more neutron rich the isotope, the easier the measurement, because neutron rich isotopes have larger weak charges and larger $A_{pv}$.

Since measurements of $R_n$ are feasible in many nuclei, one can use this freedom to choose an element that makes a very robust target.  Alternatively, PREX uses $^{208}$Pb because of its very simple nuclear structure.  The doubly magic $^{208}$Pb is an excellent closed shell nucleus where a variety of corrections, such as pairing correlations, may be small.  This allows a clean interpretation of a $R_n$ measurement in terms of bulk properties of neutron rich matter such as the equation of state \cite{alexbrown}.

However, in this paper, we find that measuring $R_n$ in $^{48}$Ca is very interesting because $^{48}$Ca is much smaller than Pb.  As a result $R_n$ can be measured faster with a higher figure of merit.  In addition, $R_n(^{48}$Ca) may be more easily related to two nucleon and three nucleon interactions, including very interesting three neutron forces, than $R_n(^{208}$Pb).    This is because $^{48}$Ca has fewer nucleons  than $^{208}$Pb and this greatly simplifies microscopic coupled cluster or no core shell model calculations.  In considering a parity violating electron scattering experiment on $^{48}$Ca, one should also consider the information that may already be available on the distribution of f 7/2 neutrons from inelastic magnetic electron scattering.  However the interpretation of this inelastic data may be somewhat model dependent because of nuclear structure effects such as core polarization. 

After measuring $R_n$, one can constrain the surface thickness of the neutron density $a_n$ with a second measurement at somewhat higher momentum transfers.  We present error estimates for measuring $a_n$ in $^{48}$Ca, $^{120}$Sn, and $^{208}$Pb.  Again, we find that $a_n$ is easier to measure in smaller nuclei.  One should study the sensitivity of the surface thickness to different features of the effective interaction to determine the possible nuclear structure information that would be available from a measurement of the surface thickness.

Finally in future work, we will present statistical error estimates for using measurements at several momentum transfers to determine the complete neutron density $\rho_n(r)$ in a model independent fashion.  This appears feasible, but difficult, for $^{48}$Ca.  However determining $\rho_n(r)$  may be extremely difficult for $^{208}$Pb.   This measured neutron density for $^{48}$Ca, combined with the previously measured charge density, will provide a very detailed picture of an atomic nucleus.

We find that parity violation experiments are feasible for a variety of nuclei and that experiments on lighter nuclei have larger figures of merit, in general.  This further motivates studies of correlations between the neutron radii of a variety of nuclei, see for example \cite{brownAPV}.  These studies should consider many nonrelativistic and relativistic effective interactions and include deformation, pairing, and other nuclear structure effects.  If these studies show that neutron radii in heavy and medium light nuclei are not strongly correlated than an additional parity experiment in a lighter nucleus such as $^{48}$Ca should provide new nuclear structure information in addition to that provided by PREX for $^{208}$Pb.

Alternatively, if neutron radii in heavy and light nuclei are strongly correlated, than our results may be even more important.  In this case one may have considerable freedom to choose the nucleus to minimize the experimental difficulties of a measurement.  For example, by following up the PREX experiment with a $^{48}$Ca measurement one may be able to significantly improve the accuracy of the determination of the neutron radius, beyond what is actually achieved in PREX.  This is because the larger figure of merit allows more statistics to be accumulated in a shorter time, while the larger parity violating asymmetry may reduce some systematic errors.  A more accurate determination of the neutron radius could be very useful to better constrain effective interactions for nuclear structure.        
\ \

This work was supported in part by DOE grant DE-FG02-87ER40365.

\vfill\eject

\end{document}